# Software Spectral Correlator for the 44-Element Ooty Radio Telescope

[1]Peeyush Prasad, C.R.Subrahmanya
*Raman Research Institute, Bangalore*

Abstract--- **A Spectral Correlator is the main component of the real time signal processing for a Radio Telescope array. The correlation of signals received at each element with every other element of the array is a classic case of an application requiring a complete graph connectivity between its data sources, as well as a very large number of simple operations to carry out the correlation. Datarates can be extremely large in order to achieve high sensitivities required for the detection of weak celestial signals. Hence, correlators are prime targets for HPC implementations. In this paper, we present the design and implementation of a massively parallel software spectral Correlator for a 44 element array. The correlator handles ~735 MB/s of incoming data from the 44 spatially distributed sources, and concurrently sustains a computational load of ~100 Gflops. We first describe how we partition the large incoming data stream into grouped datasets suited for transport over high speed serial networks, as well as ideal for processing on commodity multicore processors. An OpenMP based software correlator optimized for operation on multicore SMP systems and implemented on a set of Harpertown class dual processor machines is then presented.**

## 1. Introduction

A Multi-element Radio Telescope is a spatially spread array of antenna elements whose responses are noise-like, with responses of a pair of elements (called a baseline) required to be time aligned, dynamically calibrated and combined or correlated in real time in order to extract weak correlated components buried in the spatio-temporal correlations. These are signatures of a minute level of mutual coherence arising from weak celestial signals, and the information of interest. Thus, a Correlator is the main component of the real-time signal processing for a Radio Telescope.

In this paper, we present the design of a software spectral Correlator for a 44-element interferometric array resulting from the on-going modernization of the Ooty Radio Telescope (ORT) **[5].** Here, each antenna element is an analog phased array beam resulting from a 11.5mX30m segment of the 510m telescope with a peak bandwidth of ~40 MHz. This upgrade will lead to a 4X increase in the operating bandwidth of ORT, as well as doubling the instantaneous field of view. The Correlator itself is of a novel hybrid design and uses customized hardware to preprocess and arrange incoming data optimally, so as to assist actual correlations being formed on commodity class processors.

The ORT is capable of tracking a single sky region for 9.5 hours to achieve high signal to noise ratios. The Nyquist-sampling of the 40 MHz band for all elements amounts to 3.52 GigaSamples/Sec (GS/S), and will generate more than 50 TB of data (at 4-bit quantization) in a single observing session. The Correlator reduces these volumes to manageable levels for archival and offline processing. Here, the main operation consists of splitting each incoming datastream into required number of spectral channels using FFT, and cross-correlating the resulting spectra from every possible pair of antennas for each spectral channel to form the Cross Power Spectra (CPS). This is then accumulated to the required cadence.

For any correlator with bandwidth *B*, assuming $K.log_2K$ complex multiplies and additions per *2K* point transform, the total computation rate for *N* antennas is $N.B.log_2K$ Complex Operations per Second (COPS). Factoring the *K* complex cross multipliers and accumulators for every *N(N-1)/2* baseline leads to a total computation of $B.[N.log_2K + N(N-1)/4]$ COPS **[1].** For the case of the ORT, the 44 incoming data streams, when sampled at R MS/S, require a total real time computing of about

---

*0.99R* Gflops for spectral decomposition (512 point FFT). A computing of *1.94R* Gflops needs to be sustained for the complex cross multiplication and accumulation following the FFT. In this paper, the sampling rate R will be restricted to about 35 MS/S corresponding to a supported bandwidth of ~17 MHz. Even at this sampling rate, the real time computer platform is required to sustain over **102 Gflops** concurrently with a total input data rate of **770 MB/s** corresponding to 4-bit quantization. Since this is not possible on a single machine, a parallel approach is necessiated.

While such I/O and computing requirements used to be met by custom or ASIC-based hardware till recently, groups of commodity processors coupled with parallel processing frameworks have reached a stage where software platforms on commodity clusters can provide a more practical alternative. This also gives rise to the concept of a *Programmable Telescope* which can rapidly change its operating modes due to being software based. This approach is steadily gaining popularity, as evidenced by the recently commissioned Giant Meterwave Radio Telescope Software Backend **[4]**. Further, the current trend of many processing cores being made available in commodity processors, the even faster rate of core integration in GPUs, as well as the ability to carry out *fused multiply add* operations with hardware support on modern CPUs has led many in the Radio Astronomical community to evaluate such processors as viable correlator building blocks **[2]**.

The system described in this paper conforms to a proposed *Networked Signal Processing System* (NSPS) architecture **[3]**. The NSPS provides configurable routing of high speed data via custom switches and a buffering subsystem. It has powerful, distributed preprocessing capabilities based on commodity FPGA and high speed networking elements and allows us to partition the streaming data along one of several dimensions to enable tunability of task decomposition.

The rest of this paper is organized as follows: In section 2, we dwell upon the possible approaches to the parallel implementation of the Correlator, while we go into implementation details in section 3. We present some performace results in section 4, before concluding in section 5.

## 2. Approaches towards a parallel Correlator implementation

There are a variety of dimensions to the spatial correlation problem along which parallelizing can be carried out. Correlation can be an ideal candidate for parallelization if the entire compute load is segregated into working sets which have a certain level of independence. As noted earlier, the main tasks of a spectral correlator are the Fourier transform and the complex cross multiplication.

### *2.1 Task decomposition*

A *frequency domain* task decomposition refers the processing of a subband to a single processor. The Frequency domain can be reached via an embarassingly parallel FFT on each data stream, and subsequently, corresponding frequency channels from each sensor element can be routed to a single processor for forming all pairs of correlations for that channel. The frequency analysis may then be required to be done in hardware. This approach requires another exchange of CPS sub-band data at the end of processing for forming the per baseline full band CPS. Also, dispersion in the data path can lead to the need for a phase component to be applied to the CPS subband before accumulation in time, which can become tedious here.

In a *baseline domain* task decomposition, all time domain data from every pair of sensors is routed to a single processing element. This approach requires duplication of data, although this can conservatively be carried out via high speed commodity switches with a multicast address. The NSPS contains resources to route data to peer processors at high speeds, which makes this approach feasible.

The *time domain* task decomposition requires the collation of a continuous timeseries from all elements, before being transferred to a single processing element for processing. This workset is truly independent of every other set. Such an approach is made possible in our case by referring the data partitioning to the NSPS. This approach has the major advantage of ease of synchronizing sensor elements. The disadvantage here is the centralization of data collation, i.e., ultimately there is a single node which will contain the same set of samples from every sensor.

## 2.2 Data decomposition and the NSPS

The data partitioning depends on the task decomposition chosen. It is compounded in the Radio Interferometer by the distributed nature of sensor elements, which increases the synchronization requirement between data from remote sources. In the correlation problem, inspite of the complete graph data connectivity required, working set independence can be manipulated to any degree (and at any level of granularity of processing) by suitably manipulating the data partitioning before despatch to processors. This, thus, is the biggest advantage of having a traffic shaping network like the NSPS, as it can handle changes in correlator configuration (scalable) as well as generate working sets appropriate for general purpose processors.

The NSPS is primarily a network of FPGA based computing and buffer nodes interconnected by a high speed peer-to-peer network. It offers the flexibility of significant preprocessing of streaming data, allowing multiple computing passes on data via the large amounts of available RAM (~2GB DDR2) per node. Apart from other operations, in the current implementation it reconfigures the incoming ~770 MBps of data into frames containing 512 contiguous samples from *all* elements in a single atomic packet. Each of these packets constitutes an FFT block, giving us a basic spectral resolution of ~17KHz. Each packet is completely self contained and capable of individual existence, as it carries metadata pertaining to sample timing, datatype, number of elements etc. in its header. These packets are striped across 8 GigE links by the NSPS. To note, the computing can be matched to the data bandwidth by the addition of more computing nodes to the GigE network. Thus, the NSPS allows us to conveniently implement different data partitioning schemes.

## 3. Correlator Design and Implementation

For streaming Correlator implementations, the four main issues to contend with are:

(a) Handling of streaming input data rates by the I/O processor (b) The rearrangement of incoming data to extract maximum leverage from available compute instructions, including Fused Multiply and Add (*FMA*) instructions (c) Most effective task and data decomposition to reduce coupling between working sets, and (d) Handling scalability; increase in computing or I/O requirement due to correlator reconfiguration.

Due to the reconfiguration of the incoming data by the NSPS, our correlator implementation can apply **S**ingle **A**lgorithm **M**ultiple **D**ata parallelism via OpenMP. Hence, the ideal target for our implementation are commodity SMP nodes, which are also currently optimal from the cost per core dimension. Our implementation is tuned towards using SIMD extensions available on the target Xeon processors.

### 3.1 Workload configuration of a processing unit: I/O and compute core pair

The basic tasks needed in the Correlator implementation are shown in Fig. 1. We choose a *pair of cores* on commodity processors as our basic computing unit. One of the cores (called the I/O core) is dedicated to reading/writing of data from memory, spectral decomposition via FFT, as well as any internal memory rearrangement to ease later operations. The other core (called the compute core) is dedicated to carrying out the compute intensive cross correlation (XMAC) operation on the data. Synchronization between computing units is enforced by using OpenMP barriers by the master thread.

This situation is shown in Fig. 1., where the red blocks depict independent threads on the same core. The blue boxes, interconnected by arrows, depict the serial flow of the algorithm in a pipeline, which is synchronized by the external signal from the I/O thread on getting a set of valid frames of data. It is imperative to match the incoming data at the I/O core to the processing capabilities of the compute core. The data processing is sufficiently independent to allow increase in bandwidth or number of elements to be met by replication of the core pair alone. OpenMP thread affinity is used to tie parallel regions to the specified cores.

### 3.2 I/O considerations of a processing unit

The streaming, asymmetrical nature of our I/O needs to be balanced in the target system for maximum efficiency by utilizing architectural features of the target. In our implementation, there are 2

GigE links available on dedicated PCIe x1 lanes, which are used to bring in the data.

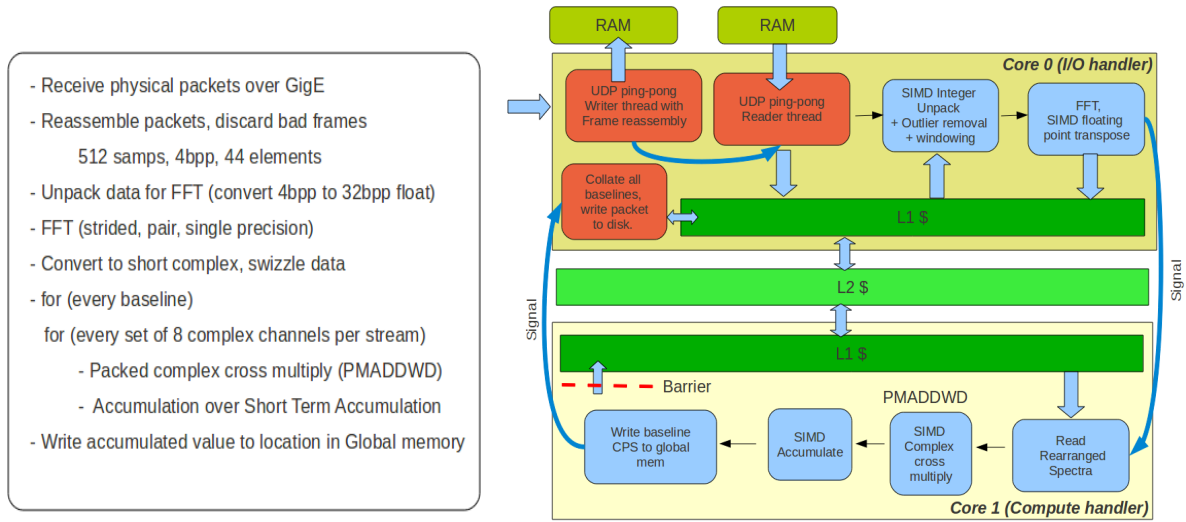

*Fig. 1:* Job partitioning between the *core pair*, a basic computing unit in our implementation

2 core pairs handle the processing of one such link, which can then be symmetrically directed towards each processor. This is possible as our target architecture *does* have features which allow load to be laid out symmetrically. The multiple processors making up the system are connected via independent high bandwidth Front Side Buses (20 GBps) to the memory controller, as shown in the figure. Incidently, our use of commodity unconnected protocols like UDP allow us to ride on the OS's I/O controller's buffering optimizations, resulting in sustaining upto 100MBps per GigE link into memory using the target system.

### *3.3  Target system description*

The software correlator has been implemented on server class machines populated with dual quad-core Xeon E5430 CPUs. Each core has 32K of I and D L1 cache while the large 6MB L2 cache per processor is used to store accumulator outputs for the ~1000 baselines. Each node is populated with 16GB of DDR2 RAM. The software environment is a 64bit Linux distribution with kernel 2.6.29. The primary compiler of choice is the gcc 4.4 compiler which implements OpenMP 3.0 specs. A mapping of our processing elements to actual hardware, as well as the distribution of workloads is presented in Fig. 2. Here, we match the incoming data rates to the computing possible with one processing unit (2 cores). Significantly, the time multiplexing carried out by the NSPS tree allows us to exploit the inherent concurrency in the target hardware via use of independent data paths, RAM regions and independent cores, which allow usage of parallel resources in the machines with very high efficiency.

## 4. Results

We report a total speedup of a factor of 42 for our optimised, multicore implementation against a single threaded reference implementation running on the same hardware for identical computing parameters. This corresponds to a ~30MSPS data handling capability per thread. For our FFT size of 512 samples, we find the Intel IPP vectorised single precision float, forward, unnormalised FFT more efficient than the most optimal option from the FFTW3 library.

The computing time is dominated by the XMAC (~75% of total load) followed by the FFT (~11%). In the absence of an integer FFT library, we find carrying out the FFT as float, a vectored conversion of complex output to short complex followed by a short complex cross multiply the most optimal approach to the problem. We also find that the XMAC is not compute, but I/O bound. This is due to the large number of complex accumulators (1 per spectral channel per baseline) which currently overflow L1 cache by a factor of ~25 and cause a large number of L1 misses. We have found that the optimal FFT output data arrangement for the XMAC to match spatial locality and available SIMD instructions is a memory grouping of 4 consecutive timeslices of each antenna's FFT spectral channel as short complex. This, combined with the PHADD/SUB horizontal addition/subtraction instructions available in SSSE3 result in the fastest code (~6GMACS/s per core, measured on our hardware), keeping in mind memory access issues. We are working on incorporating the FFT within the NSPS hardware, followed by this optimal grouping of spectral data by the NSPS itself. The loss of precision while converting to short int from float is compensated by prescaling, also available in the Xeon ISA, but due to the statistical nature of the estimation problem, this is not a crippling problem.

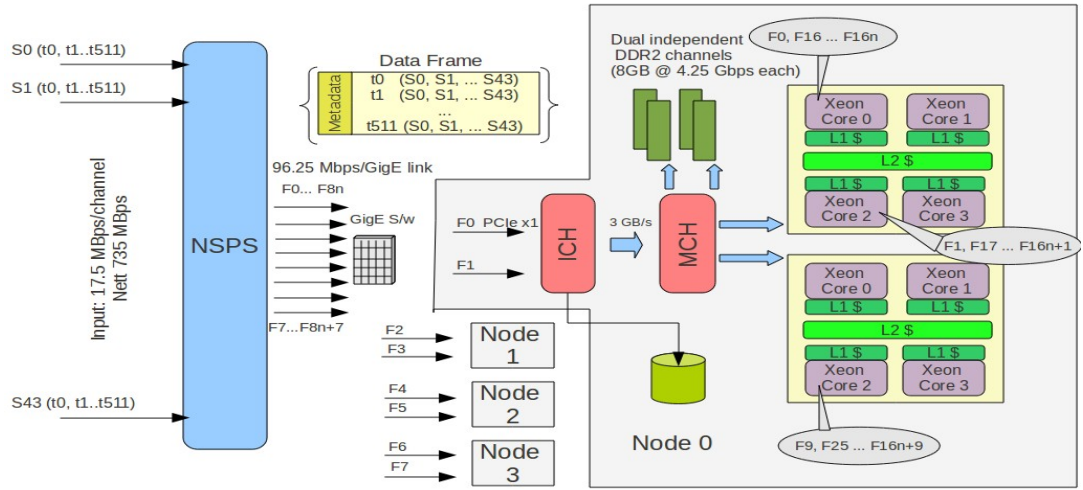

*Fig. 2:* Correlator implementation on Xeon quad core, dual processor machines showing the data distribution which exploits the architectural symmetry

## 5. Conclusions

Explicit parallelism in correlator designs is required to take full advantage of the increasing number of compute cores being offered by commodity processing. The NSPS is a flexible preprocessing architecture catering to the latency critical section of processing in an interferometric array. Due to its large available link bandwidths, a highly configurable telescope backend can be formed with the flexibility of software DSP. It carries out the job of reorganizing data to enable explicit parallelism as well as tune preprocessed data to suit a target implementation architecture.

In this paper, we have presented the detailed design of a parallel software correlator suitable for handling the real time processing of a 44 element array. The correlator has been implemented using 4 nodes with features as described earlier. This amount of computing is comfortably catering to the telescope requirements. The correlator depends on custom hardware in the form of the NSPS to reconfigure streaming data at high speeds. We advocate forming time domain multiplexed data containing samples from all elements as the basic compute data frame. We have also established a pair of cores as a viable processing entity for the compute data frame, and shown that a group of such core pairs coupled to the NSPS can effectively partition the problem and solve it.